\documentclass[aps,prl,twocolumn,showpacs,superscriptaddress,groupedaddress]{revtex4}  
\usepackage{graphicx}  
\usepackage{epstopdf}
\usepackage{dcolumn}   
\usepackage{bm}        
\usepackage{amssymb}   
\usepackage{natbib}
\usepackage{hyperref}

\begin{document}

\title{A Low Reynolds Predator}

\author{Mehran Ebrahimian\footnote{Current Address: Sharif University of Technology, Graduate School of Management and Economics, P.O. Box 11155-8638, Tehran, Iran.}} 

\author{Mohammad Reza Ejtehadi}
\email{ejtehadi@sharif.edu}
\affiliation{Sharif University of Technology, Department of Physics, P.O. Box 11155-9161, Tehran, Iran.
}

\date{\today}

\begin{abstract}
Here we introduce a two-dimensional (2D) low-Reynolds swimmer and discuss the motion of the swimmer both in noise-free and stochastic regimes. Three spheres, linked by extensible arms, in a plane form the triangle body of micro-swimmer. Expansion and contraction of two-stated linkers in appropriate order causes both translational and rotational movement of the swimmer. It is shown that the motion of the swimmer could be controlled from a rotor to a directed linear swimmer depend on the sequence of the linkers activities. A few amount of noise in the the rhythm of the cycles introduces an interesting "random arc", while in the case of completely stochastic activity of the linkers the swimmer goes to a regular diffusive regime. Moreover, we show that in response to an external source of disturbance, the swimmer may approach or escape the source and shows interesting chemotaxis behavior. It is also shown that the model swimmer can easily be generalized to three dimensions or more complicated geometries.
\end{abstract}


\maketitle

The problem of swimming at low Reynolds number is relevant to life in micro-scale \cite{1}. But swimming is not enough as the living cells should look for what they need. Although at a first glance the movement of a single-cell microorganism seems random, the ability to approach food resources or escape from hazards (chemotaxis) is essential for life \cite{2,3}. Many simple and self-propelled microswimmers at low Reynolds numbers have been suggested \cite{4,5,6,7}.  To find how such swimmers can simulate a chemotaxis process, one should investigate the problem in more than one dimension. 

Recently, Najafi and Golestanian have introduced a simple swimmer and showed that it works well at low Reynolds number \cite{8}. The swimmer consists of three solid beads, contacting with two extensible tiny linkers in a line. The linkers change their length in a non-reciprocal but cyclic way. Because of the screening effect the viscous friction, applied to the beads by the fluid, depends not only on their speed, but also on the distance between the beads. Thus the swimmer displacements do not cancel each others in a full period, and it swims. Here we introduce a 2D variant of this low-Reynolds number swimmer. It is modeled again with three solid beads but connected together by three arms forming a triangle. The arms have negligible thickness, but they can change their length which is the source of the swimmer dynamics. Considering the symmetry of the system, it is expected that the swimmer only moves in the plane of the triangle (no dynamical symmetry breaking to let swimmer escape from the plane). Thus we call it a 2D swimmer. There have been other attempts to introduce 2D swimmers \cite{5,6,7,9}. However, we show that our model not only could simply generalizable to three dimensions, it also shows chemotaxis behavior.

\begin{figure}
\includegraphics[scale=0.32]{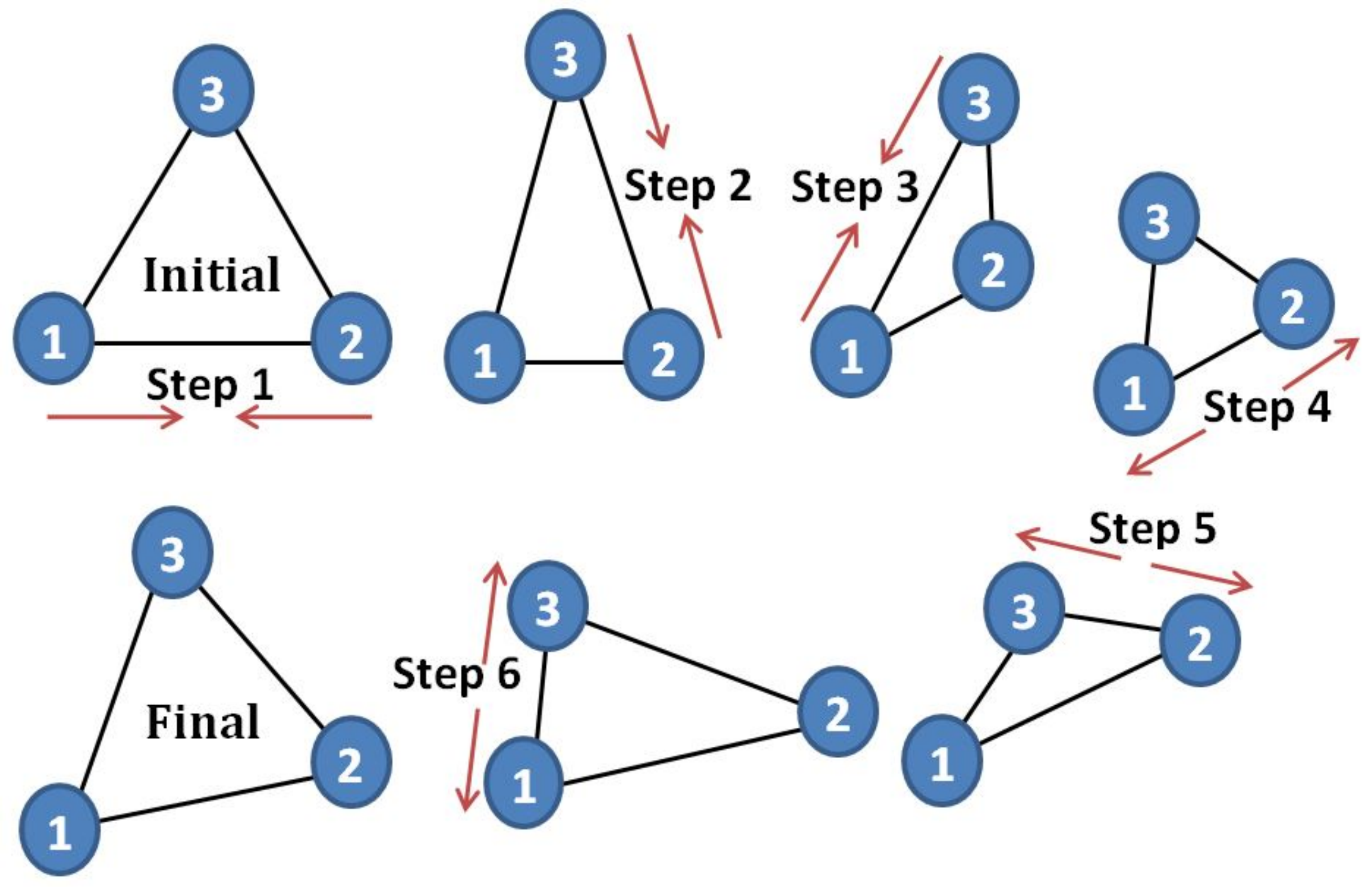}
\caption{\label{fig:1} A full regular cycle started by contraction of one of the arms (e.g. arm 1-2 here) while the other arms follow it. After half cycle they revert to their relaxed states in the same order. The last snapshot compares displacement of the swimmer after a full cycle with its initial position. }
\end{figure}

In the moving model, any linker may reduce its length from $L$ to $(1-\epsilon)L$ with a speed $W$ or restore its original length with the same speed. A full period of the cycle consists of 6 steps as is shown in Fig.~\ref{fig:1}. Starting from a relaxed situation, in half of one cycle the linkers shrink in turn and then they relax in the next half cycle in the same order. After a full cycle the body returns to its initial configuration, but with a net displacement of the center of mass (COM) and a net rotation around it, because of the hydrodynamic interactions between the balls.

For low Reynolds numbers, the nonlinear term of Navier-Stokes equation is negligible and the equation that describes the hydrodynamics of the swimmer in an incompressible flow condition i.e. $\nabla.\textbf{u}=0$, is $\mu {\nabla}^2 \textbf{u}-\nabla p=0$, where $p$ and $\textbf{u}$ represent the pressure and velocity fields and $\mu$ denotes the fluid viscosity. Zero velocity at far distances, and no-slip boundary condition on each sphere, are assumed and the dynamical effects of the linkers on the swimmer's motion are neglected. Following \cite{8} the linear form of the equations let us to find a relation between velocity of the ith sphere ($\textbf{V}_{i}$) and the force applied to it ($\textbf{F}_{i}$) as $\textbf{V}_{i}=\sum^{3}_{j=1}H_{ij}\textbf{F}_{j}$, where ${H}_{ij}$ is the symmetric Oseen tensor that depends on the geometry \cite{10}. By considering Newton's laws on conservation of linear and angular momentums ($\sum\textbf{F}_{i}=0$ and $\sum \textbf{r}_i \times \textbf{F}_i=0$), the set of equations are complete and can be treated numerically. The limited number of possible configurations for the body let us solve the equations for any geometry once and save them to perform our simulations.

\begin{figure}
\includegraphics[scale=0.25]{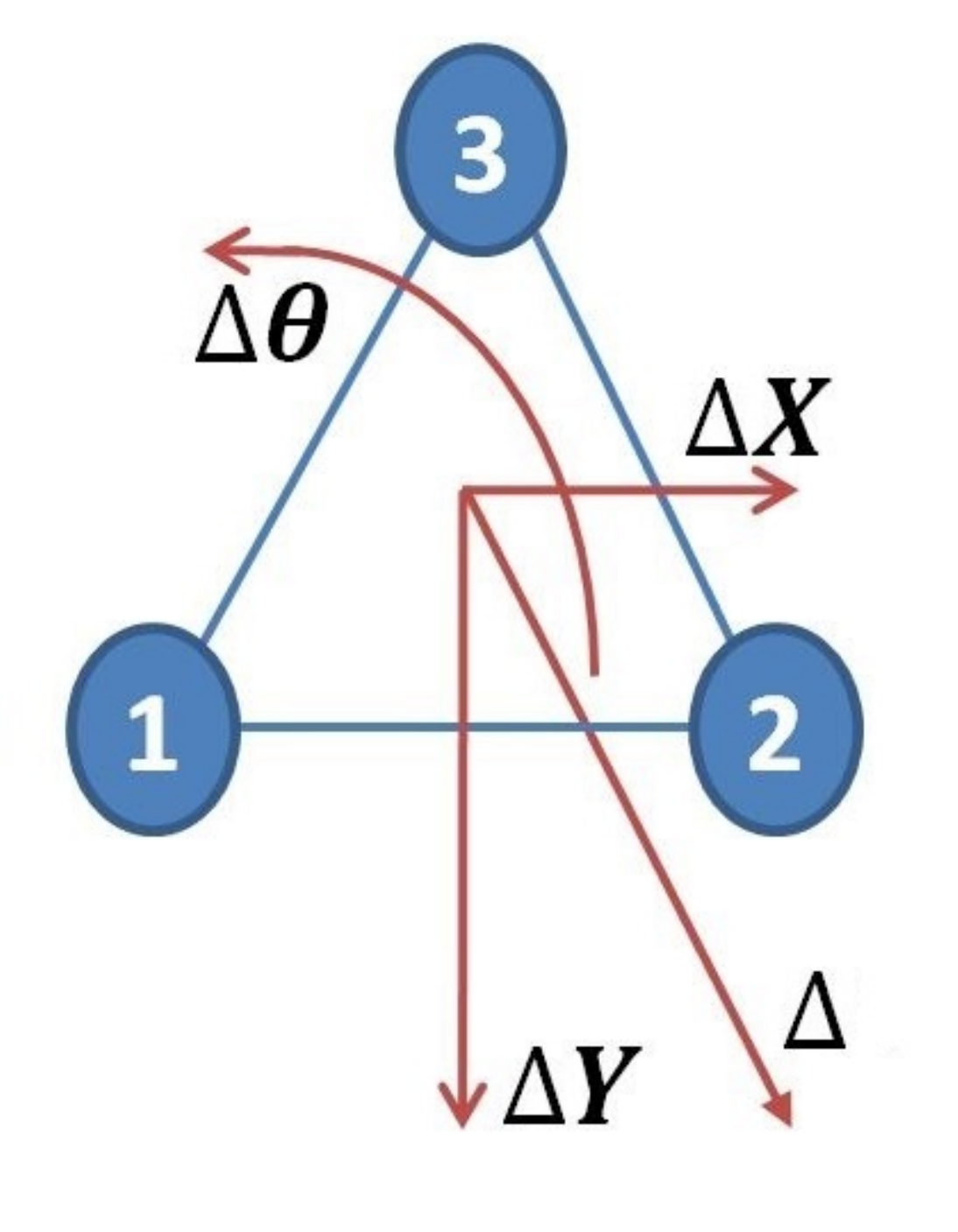}
\includegraphics[scale=0.41]{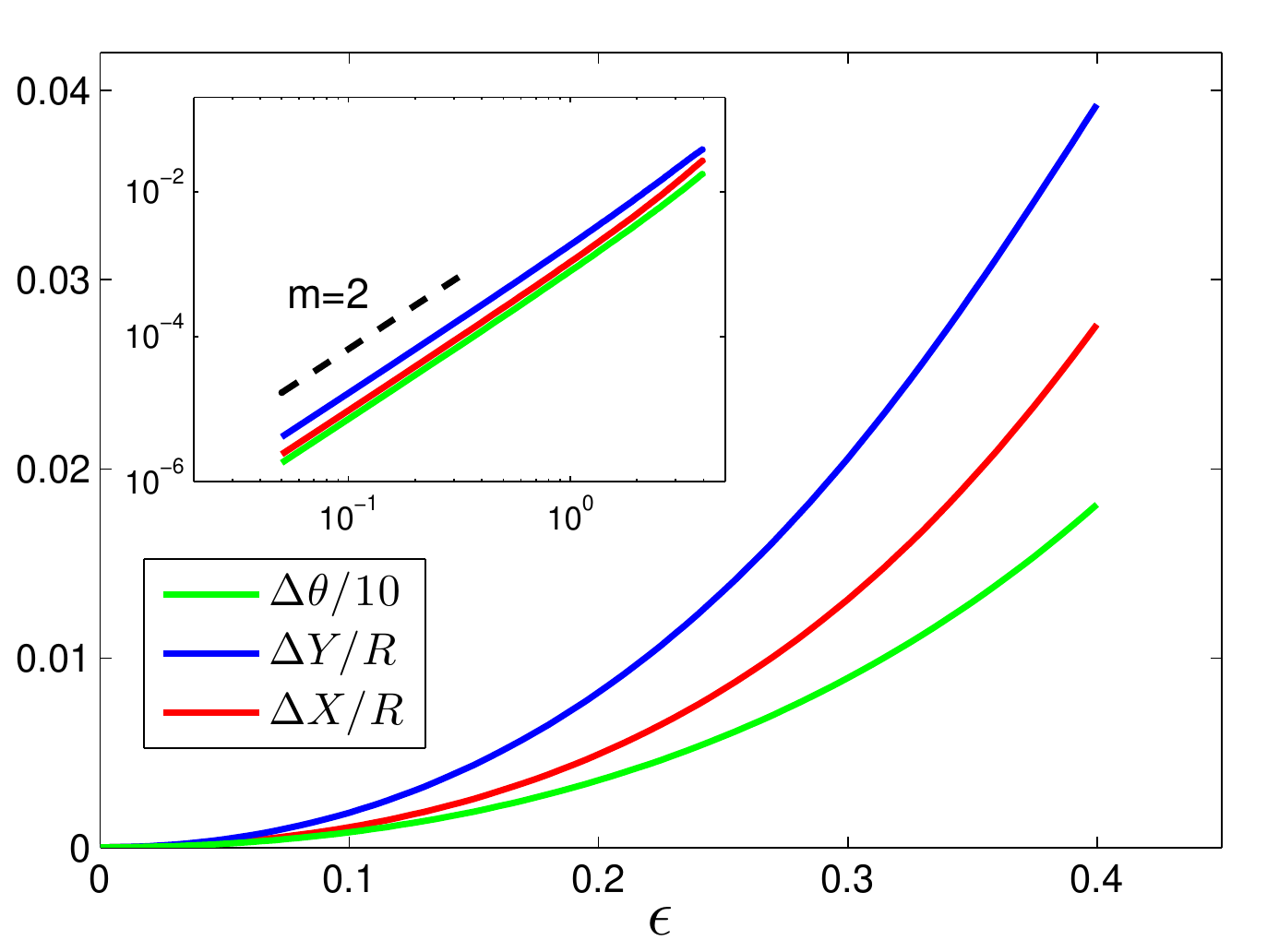}
\caption{\label{fig:2} After a full cycle, the body both rotates and moves. The movement, $\vec{\Delta}$, can be decomposed to the component parallel, $\Delta X$, and perpendicular, $\Delta Y$, to the initial direction of the starter linker. The displacements and rotations are shown for relaxed arm length $L=10 R$ as a function of contraction, $\epsilon$. Inset shows the displacements in log-log plot which indicated that they are second order with respect to $\epsilon$. The triangular geometry of the swimmer restricts $\epsilon$ to be not greater than $1/2$  (see intermediate steps in Fig.~\ref{fig:1}).
}
\end{figure}

Total displacement and rotation of the swimmer for a full counterclockwise cycle, are shown in Fig.~\ref{fig:2} as a function of contraction ratio, $\epsilon$, for some given parameters of the model. The radius of the spheres, $R$, is taken as the units of length and time. If the cycle starts from any other linker the final displacement can be found with a proper rotation of $120^{\circ}$. By considering all variations, there are only 6 possibilities to characterize a complete cycle. 
The movement of balls in each step is propertional to $\epsilon$. However the movement due to the contraction of a linker is not canceled by its expansion in further steps, because of different positions of other balls, which are also of the first order of $\epsilon$. Therefore, one can expect that the total displacement is second order with respect to the contraction ratio, $\epsilon$.

If the swimmer keeps continuing its regular cycles, displacements in the end of the cycles are the same with just a constant rotation. Hence the swimmer’s COM completes a circular path of radius $\rho=|\vec{\Delta}| / \Delta \theta$ after $2\pi / \Delta \theta$ full cycles. The radius of the rotation is smaller than $R$  for possible values of $\epsilon$, and much less than the size of the swimmer, $L$ (Fig.~\ref{fig:2}); hence in this mechanism the swimmer rotates almost in place and acts as a fixed rotator.

\begin{figure}
\includegraphics[scale=0.45]{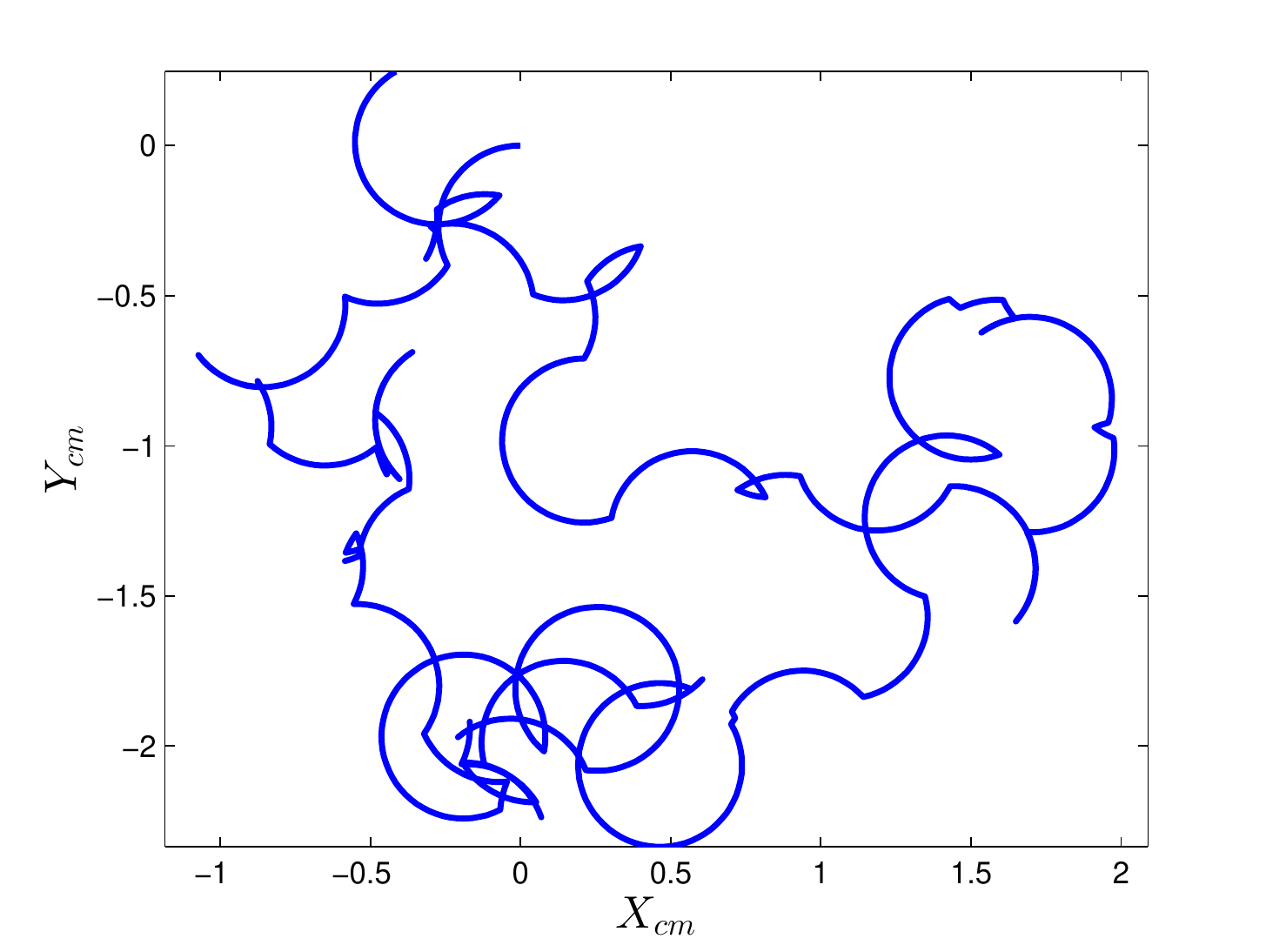}
\caption{\label{fig:3} A typical path of the swimmer in 500 cycles for $L=10 R$ and $\epsilon=0.3$ which is perturbed by the probability $p=0.1$. The unit of length is taken to be $R$. All the arcs have the same radius. The kinks are $120^\circ$ because of the triangular symmetry of the swimmer. The arc lengths are random and have an exponential distribution. }
\end{figure}

However, it is quite probable that sources of randomness, e.g. thermal or chemical fluctuations, perturb our swimmer's regular cycles. For example we suppose that in the end of any cycle the swimmer continues this cycle with a probability $(1-p)$ or start a new cycle with probability $p$ which is different from the old one either in the cycle direction or in the launching arm. Therefore, the swimmer follows circular arcs which are kinked in points of perturbation. This introduces a very interesting 2D random walk which is composed from random arcs with $120^{\circ}$ kinks in between (see Fig.~\ref{fig:3}). 

For $p\ll 1$ we can use continuum approximation by introducing the rate of direction change $\eta = p / \Delta \theta$ that is the probability rate (per arc angle) for occurring perturbation during the motion. So the arc angle has probability distribution
\begin{eqnarray}
P(\Delta \alpha_n)=\eta e^{(-\eta \Delta \alpha _n)}
\label{eq:1}
\end{eqnarray}
where $\alpha _n$ is defined as angle between tangent to the path of swimmer and a fixed direction in lab frame, and $\Delta \alpha _n$ is the arc angle of the nth arc. Simple geometry gives step length in the nth step as $l_n=2\rho \sin{\Delta \alpha_n /2}$, where $\rho=\Delta R/\Delta \theta$ is the radius of the arcs. All the arcs have the same probability distribute so $\langle{l_n}^2\rangle$ is not a function of n and we call it $l^2$. Using Eq.~(\ref{eq:1}) we have
\begin{eqnarray}
l^2=4\rho^2 \eta \int_0^\infty \sin^2(\alpha/2) e^{(-\eta \alpha)} d\alpha.
\label{eq:2}
\end{eqnarray}

Moreover, the length of each arc is independent of the state of the body in the beginning of nth arc, in particular its initial direction, so it can be conclude that $\langle r_n^2 \rangle =nl^2$, where $r_n$ denotes the position of COM in 2D space after n steps. Calculating $l^2$ from Eq.~\ref{eq:2} results in the following equation for mean square of displacemnet:
\begin{eqnarray}
\langle r_n^2 \rangle =\frac{2 \rho^2n}{1+\eta^2}.
\label{eq:3}
\end{eqnarray}
It should be noted that $n$, the number of arcs, is related to the rate and angular velocity of the walker, $\omega=\Delta \theta /\tau$, by $n=\eta \omega t$, where $t$ is time. Substituting this in Eq.~(\ref{eq:3}) leads us to the follow equation for 2D diffusion coefficient,
\begin{eqnarray}
D=\frac{\eta \omega \rho^2}{2(\eta^2+1)}.
\label{eq:4}
\end{eqnarray}
As one expects for very small values of $p$, which is small values of $\eta$, the diffusion coefficient vanishes. That is because the swimmer most of the time acts like a fixed rotator and does not move far.

In the presented swimming model, fixed rotator moves just as a result of introducing perturbation. However, minor adaptation in the algorithm of contraction/expansion may make a linear swimmer model without requiring any randomness. For a clockwise cycle, it is trivial by symmetry that $\Delta X$ and $\Delta \theta$ change their sign, with respect to the movements of counterclockwise cycle starting from the same arm (see Fig.~\ref{fig:2}). If the swimmer alternatively switches the direction of its cycles, we can have a pure linear swimmer. In this way after a pair of cycles, total rotation will be zero and COM moves approximately $2 \Delta Y$, so the velocity of linear swimmer $v=\Delta Y/\tau$.

Introducing noises in the linear mover results in a much more familiar random walk with $120^\circ$ kinks and straight steps. As in the previous case, we suppose a probability $p$ per cycle for change in the swimmer direction which results in $D=v^2 \tau/2 p^2 $, where $v=\Delta Y/\tau$ is the velocity of straight moving. In this case the diffusion coefficient is a descending function of $p$ in contrast to the first mechanism. The reason is that, whereas with the first mechanism the swimmer almost rotates in place and randomness helps it to move, in the later mechanism the swimmer moves forward with a constant velocity $v$ and randomness perturbs its directed motion.

More primitive organism may have fully random swimming. Increased strength of the randomness, finally, destroys the defined cycles in linkers’ dynamics. In a given time step each of the arms of the swimmer are either close or open. Then any randomly chosen arm does change its state in the next step. Thus, the swimmer's movement (both translational and rotational) depend both on its current configuration (8 possibilities) and the selected arm (3 choices). Geometrical symmetries reduce 24 possibilities to only 8 distinct ones. The swimmer moves randomly on the plane and does a fully diffusive random walk with a mean displacement per step that can be obtained by averaging over all 8 possibilities. 

One may expect that the diffusion coefficient of such a random walk is determined simply by considering the average step size. But since the steps are not completely uncorrelated this is not the case. An arm can act only in opposite of its previous move (there are only two states). Because of the reversible nature of the swimming in low Reynolds numbers, such reversed moves \textit{almost} cancel out each other and reduce the mobility by a few orders of magnitude. The diffusion constant which is obtained by averaging MSD over three hundred realizations for a swimmer with $L=10$ and $\epsilon=0.3$  is about $10^{-4}$. If an arm cannot be selected in two successive steps (eliminating purely de-constructive steps) the mobility increases by a factor of almost three. This assumption is reasonable if we think of the steps as the action of a micro- (or nano-) machine sitting on the arms, and if we consider a natural relaxation time for that machine.

So far we considered the situations in which the arms selection rule is homogeneous and space-independent. Then it is trivial that the average displacement of the swimmer as like as any other homogeneous random walker is zero. What about the situations in which micro machines, that are responsible for the change of length of the linkers, are sensitive to concentration of chemicals in their environment? For instance, we can assume the presence of some chemical nearby linkers, that affect chemical equilibrium according to Le Chatelier's principle: the probability of staying in the close state for linkers, as mechanochemical enzymes, may be depended on the concentration of chemicals nearby that linker, which bonding with the linker results in expansion. In the case of space varying concentration of activator chemicals the transition rate between relaxed and contracted states of body arms is not symmetric, so the swimmer may be drifted to a particular high/low concentration zone. 

\begin{figure}
\includegraphics[scale=0.56]{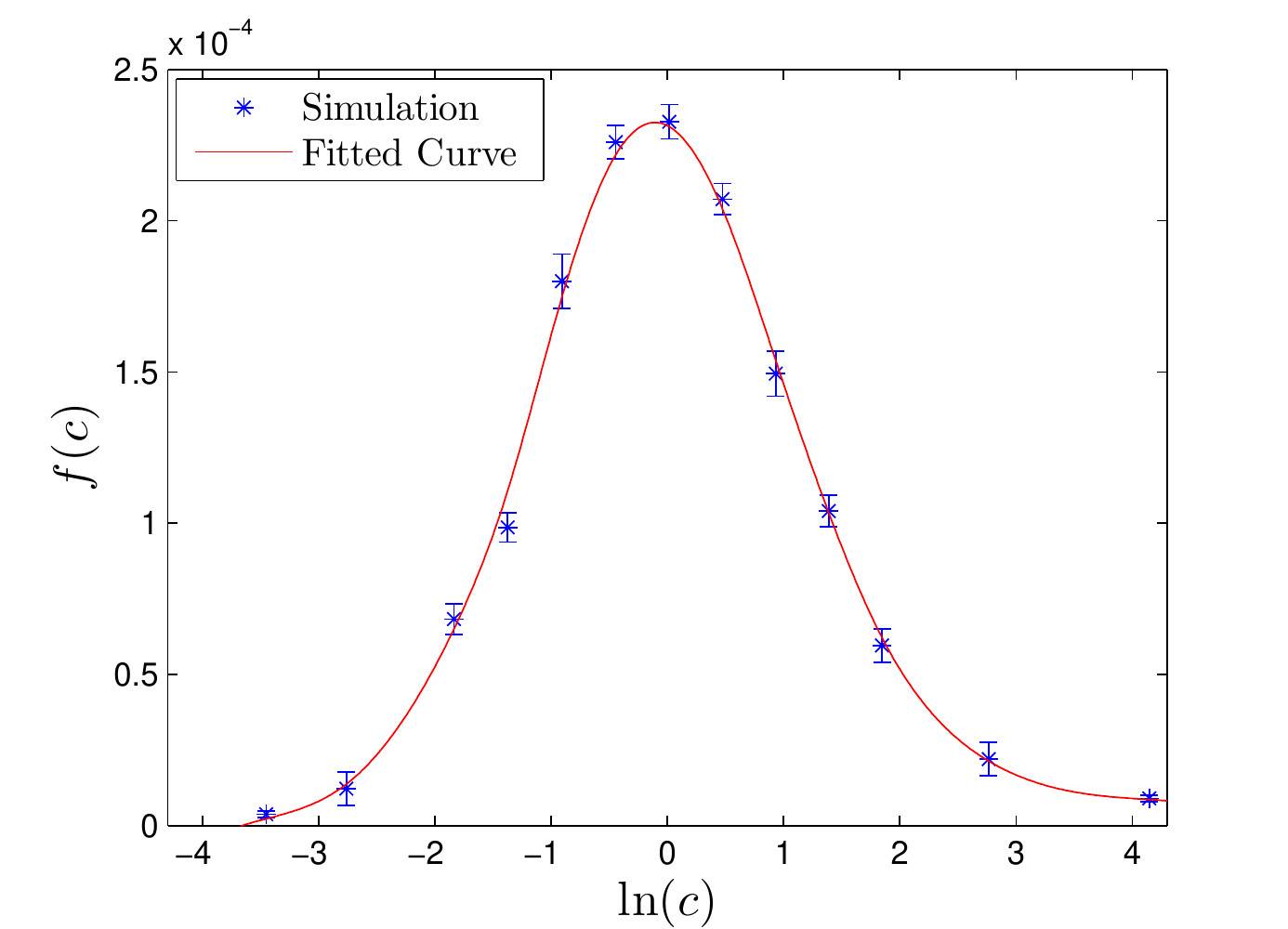}
\caption{\label{fig:6} The average drift velocity response of swimmer to variation slope of space-dependent concentration of chemical activators, as a function of concentration. For $c=1$, expected time of being in close or open state for each arm is equal. Numerical results are based on enough number/time of simulations to reach precise average values.}
\end{figure}

In the first order of expansion with respect to concentration variation, the drift velocity can be written in the form $\textbf{u}=f(c)\nabla \log (c)$, where $c$ is the concentration of chemical activators and $f(c)$, determines the slope of drift with respect to environment variations. Fig.~\ref{fig:6} plots the simulation results for $f(c)$. The unit of $f(c)$ is $R^2/\tau_{o}$, where $R$ is the radius of spheres and $\tau_{o}$ is expected time of an arm to stay relaxed. In the limit cases of $c \ll 1$ or $c \gg 1$, one state is highly preferred by all the arms, so the body reshapes very fast to fully close or open configuration. After that, each stochastic change to other state for a given arm is quickly reversed, because of tendency to being in a particular state. Hence, the consecutive state changes cancel out each other, because of reversible feature of micro world, and movement ability fails.

In the proposed model, the presence of chemicals adopts the preference of arms to be in their relaxed state. As a result, on average the swimmer moves towards the high concentration area, and it escapes low concentration area. Thus the swimmer shows a perfect chemotaxis behavior. It approaches food resources when the closest arm to the food is more probable to revert to the restored state, and it escapes when the closest arm to the hazards prefers to be in the shortened/relaxed state because of release/consumption of chemical stimulator by foods/dangerous objects. 

To demonstrate this effect we introduce a low-Reynolds predator-prey system. A tiny prey is swimming in the media. We assume the hydrodynamic effect of the prey on the motion of the predator is negligible, but its metabolism somehow changes the chemistry of its environment in a way that our swimmer's (predator's) arms prefer to be in the relaxed state when the predator is close enough (to sense the chemistry).

\begin{figure}
\includegraphics[scale=0.53]{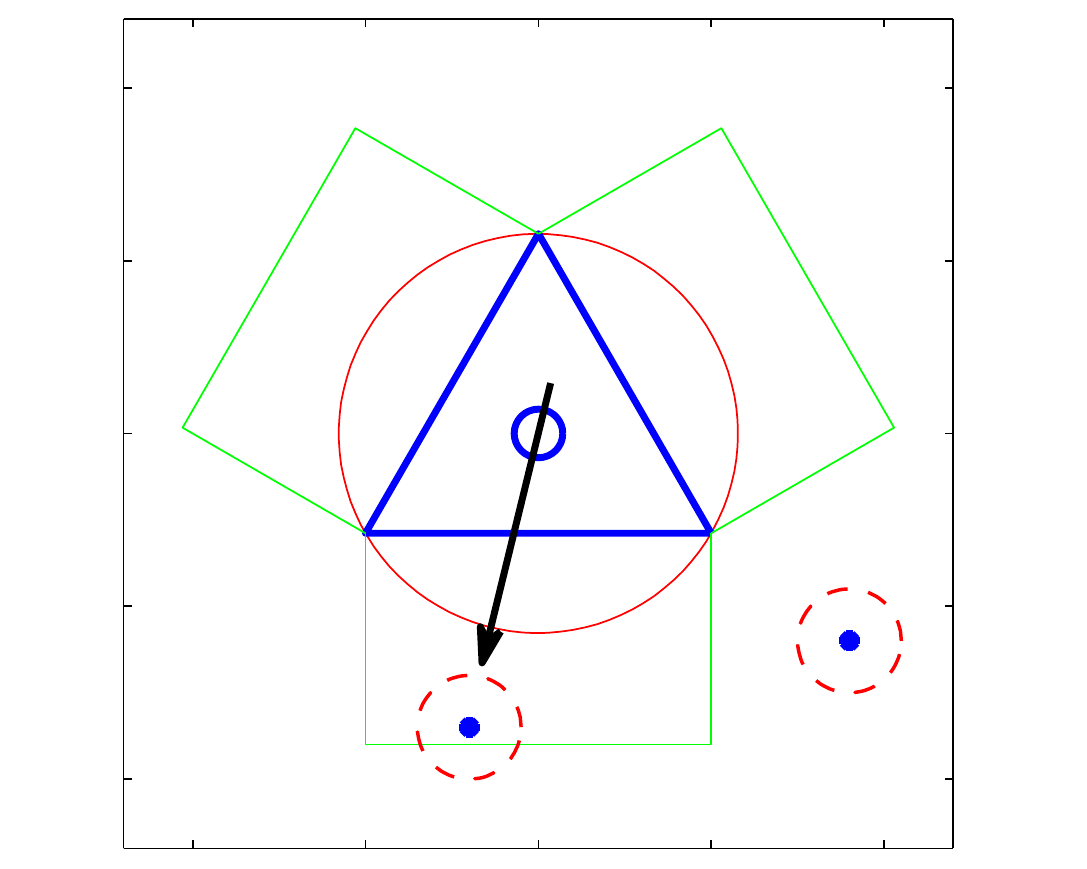}
\caption{\label{fig:5} To introduce the predator it is supposed that the prey is a source of chemicals which affect the linkers' dynamics. Without lack of generality, we simplified the interaction to a step potential and a rectangular zone is introduced for any arm and it is supposed that the linker transition rates are affected by the prey if it wanders in this neighborhood.}
\end{figure}

Fig.~\ref{fig:5} shows the geometry of our model predator and prey. The large red and small blue circles indicate the size and COM of predator, respectively. The arms of the predator swimmer are presented by blue triangles and for each arm the green rectangular area represents the area where we have assumed the state (dynamic) of the arm is affected by the presence of the prey, which are indicated by dashed red circles with blue points in the center. When the predator is blind to preys it randomly swims as described above. But if by chance a prey wanders into one of the green rectangles, it affects the arms selection rule and it is more probable that the corresponding arm reverts to the relaxed state, if it is in shorten state. With just this simple rule we see that predator approaches the prey and catches it.

The movies in \href{http://softmatter.cscm.ir/Low-Reynolds-Predator/index.htm}{supplementary materials} show how the swimmer chases its prey. In movie \textbf{s1} the swimmer is chasing an escaping animal. The smaller body (prey) is doing a simple random walk and only avoids the predator with a simple hard core repulsive potential. To demonstrate the chasing ability of the predator, in another scenario the prey is helped to escape from the predator in lateral routes (movie \textbf{s2}).

Finally, it is noticeable that extending the predator swimmer model to three dimensions is so straightforward. The simplest geometry is a tetrahedron of four spheres, connected together by 6 arms. Again, each arm can change the length as in 2D model. In any step one arm changes its state with the constant speed $W$. The equations of motions are similar to 2D case and we should only solve a bigger system of equations to find the spheres displacements. The translational and rotational displacements corresponding to this motion are smaller but still in the same order as the 2D swimmer. That is because the screening effect is weaker in 3D case. Again, if we consider asymmetry in the rates of transitions, the chemotaxis effect on the motion of the swimmer is observed and can help it to move toward (or escape from) the sources of perturbations in three dimensional space.

We would like to thank Ali Najafi for very usefull comments and his valuable hints, Ramin Golestanian for his fruitful discussions and Nader Heydari for reading the draft manuscript and his valuable language comments.

\bibliographystyle{apsrev4-1}
\bibliography{ref}

\end{document}